# ANOTHER CLASSICAL ANALOG OF GROVER'S QUANTUM SEARCH

## Shayak Bhattacharjee

Department of Physics,
Indian Institute of Technology Kanpur,
NH-91, Kalyanpur,
Kanpur – 208016,
Uttar Pradesh, INDIA

*    *    *    *    *

## ABSTRACT

In this work I describe a classical analog of Grover's quantum algorithm for searching an unsorted database. I consider a game of roulette in which the wheel has certain extra features to mimic the quantum system. It turns out that $O(N)$ steps are required in the classical case, while the answer immediately drops to $O(N^{1/2})$ when the quantum features are incorporated. This answers the question posed by Grover himself as to whether there is any succinct physical argument describing the success of the algorithm. The model also acts as a pedagogical tool to clarify the concepts behind Grover's algorithm.

*    *    *    *    *

The reason why quantum computation is one of the hottest topics in today's research is because quantum computers can be orders of magnitude faster at certain operations than their classical counterparts. There are two basic components to designing fast quantum computers – (a) how must the two-level systems and the unitary operations etc. be practically realized, and (b) given a quantum computer, in what manner must it be operated so that the desired tasks are indeed accomplished in record time. The second of these questions is addressed by quantum algorithms. Today there are two quantum algorithms which are known to be greatly faster than the corresponding classical procedures. One of these is for factorizing an integer [1] and was invented by Shor. The other is for searching a database [2] and was invented by Lov Kumar Grover.

Grover's pioneering paper of 1997 is based mostly on the mathematical aspects of the algorithm with the physical arguments, though the backbone of the paper, playing a secondary role. Years later in a pedagogical publication, Grover tried to highlight the physical features of the algorithm. By his own account this was not a very successful venture as the paper ends with the lines, "What is the reason that one would expect that a quantum mechanical scheme could accomplish the search in $O(N^{1/2})$ steps ? It would be insightful to have a simple two-line argument for this....."

One of the reasons why it is often difficult to gain insight into a quantum problem is because the systems encountered there cannot be seen with our own eyes, and are hence unfamiliar. Classical systems on the other hand are part of our daily experience, hence many people have strong intuitive feel for such systems. It is logic of this form which has driven several authors to devise classical analogs of Grover's algorithm. One of these authors [4] is Grover himself. He has considered a system where there are $N$ pendula, one slightly shorter than the others. It turns out that if the pendula are suitably coupled, then $O(N^{1/2})$ oscillations are required to identify the odd man out. Grover claims that this is identical to the result of his quantum algorithm. This analogy however is flawed because $O(N^{1/2})$ oscillations of $N$ pendula do not amount to $O(N^{1/2})$ steps but rather $O(N^{3/2})$ steps of run of the computer. Moreover, for the algorithm to work, there must be precise tuning and detuning relationships between the various pendula, hence the system appears contrived and is of limited interest. A second analogy was proposed by Zhang and Lu [5], who consider a system where repeated one-dimensional collisions are engineered between two balls. It is seen that the entire kinetic energy of the system can be transferred preferentially to one of the balls, just as the amplitude of the target state is preferentially increased in Grover's algorithm. This analogy too is of limited utility – for one, there is nothing being searched for in this procedure and for another, a single collision between balls of nearly equal mass can transfer almost all the energy from one ball to the other. In this work I try to propose a classical representation of Grover's algorithm which overcomes the deficiencies of the previous attempts. To achieve this, I stick as close to Grover's original logic as possible, only replacing the quantum system with a roulette wheel. It turns out that consideration of this system also gives a possible answer to Grover's query which I mentioned in the previous paragraph.

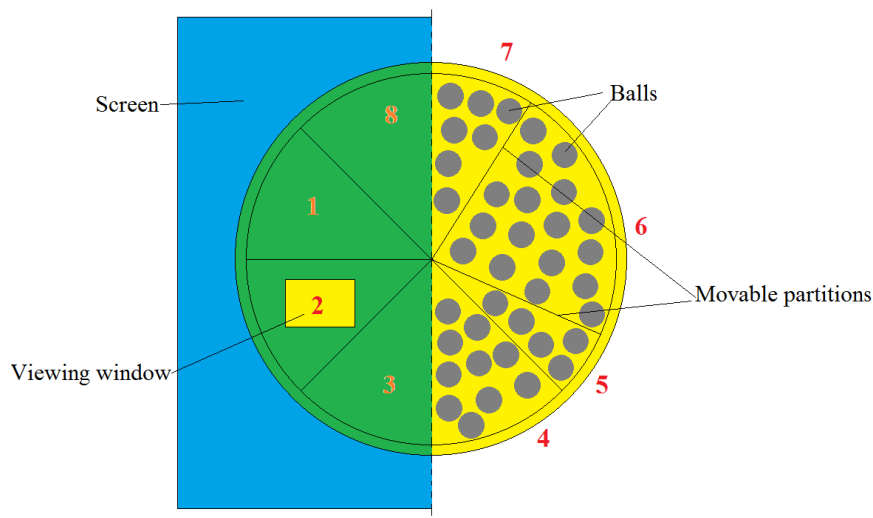

*Figure 1 : A setup for playing roulette. The machine (yellow) is hidden from view by the screen shown in blue and truncated at the halfway point for clarity. The viewing window is also seen. The machine is divided in N=$2^3$ labelled angular sectors. The partitions between these sectors are movable and the space occupied by any sector depends upon the number of balls inside it.*

Consider a roulette wheel (Fig. 1) partitioned into $N=2^n$ angular sectors (this condition has been chosen only to mimic the original $n$-qubit system, and the logic works for any sufficiently large $N$). Each sector is labelled with a unique number from 1 to $N$. The partitions are movable so that the area allotted to each number can be changed (for instance by a crooked casino master). Each compartment of the roulette is filled with innumerable tiny balls, the area occupied by the sector being proportional to the number of balls inside it (for this step the balls can be viewed as gas molecules in equilibrium). Adding or subtracting the balls in any sector automatically increases or decreases its relative weightage. So that the gamblers may not cry foul play, the roulette is hidden behind an opaque screen. This screen has one small viewing window – the outcome of a throw of the roulette is the number appearing in this window when the wheel comes to halt. Now suppose I am given the roulette in the unbiased state (fair casino master – equal area for all sectors) and am told that I get the jackpot if one and only one particular number appears in the window, and lose my money in all other cases. My task is to determine (in as few steps as possible – profits must be maximized) which is in fact the magic number. As a trade-off between speed and accuracy I will be given the leeway of not having to determine the jackpot number exactly but only with a macroscopically high probability (say 1/2).

With the classical roulette, my solution is well known and fairly straightforward. I must keep throwing the roulette until I get the jackpot, and then I must note down the magic number. On an average, I will need about as many throws as there are numbers on the roulette i.e. I will find the magic number in $O(N)$ tries. Now because the roulette is in fact a representation of a quantum system, I allow certain operations which are not practical with ordinary roulettes. In other words, I invoke an army of ants which will access the roulette through the viewing window and internally manipulate the balls as per a programmed routine. Of course the ants cannot communicate the results of their manipulations to me – I am no zoolinguist.

Now, I stipulate the following routine for the ants. Suppose the target compartment (i.e. the sector which fetches the jackpot) is given a unique internal label which only ants can understand. Then the ants perform the following two operations.

1. Go to the jackpot sector and remove all the balls from it.
2. (i) Calculate the average number of balls per sector (let us say it is $A$).
   (ii) [repeat over all sectors] If the number of balls in any sector is $A \pm B$ then change the number of balls in that sector to $A \mp B$, where positive and negative signs correspond. This operation has been termed by Grover as 'inversion about average' which is a clear enough label.
   (iii) Replace all the balls which have been removed in step 1.

Now let us see what will be the effect of performing these operations, starting from a uniform distribution of balls. After the first operation, the jackpot sector is devoid of balls while all the other sectors have equal number of balls. In the limit that $n$ is large (and hence $N$ is huge) the average number of balls per sector will remain at its starting value. Now when the second operation passes over the sectors, it will do nothing about the filled sectors (which are already at average value) but will stop at the empty sector. Since this is below average to begin with, it will become above average after the inversion (twice the average to be mathematically precise). Then the removed balls will also be restored to make the total stand at thrice the average. Since the angular span of each sector is proportional to the number of balls inside it, the jackpot sector will now occupy a little more area than all the others. If I run the two operations again, the jackpot sector will become larger still, and this will continue with each iteration of the operations. Hence at the end of multiple iterations, the jackpot sector will be far larger than all the other sectors. If I now recall the ants army and throw the roulette, there will be a large probability of its coming to rest with the jackpot sector neatly positioned under the viewing window.

A fine point in this reasoning is that the number of balls inside the roulette is increasing over time. If we think of the balls as gas molecules then this picture is acceptable – we are filling up the roulette with compressed air. Otherwise we can always imagine some kind of normalization over all sectors to maintain constant density of the roulette – this is a gedanken experiment after all and this implementational detail is not of the highest importance.

Now a question which *is* of highest importance – how many iterations of the two operations will be required ? Suppose that at the start of the process there are $\alpha$ balls in each of the non-jackpot sectors. Because the jackpot sector is cleared each time before carrying out the inversion, the number of balls in the other sectors will not change but that in the jackpot sector will increase by $2\alpha$ after each iteration of Operation 2. Now a ball number of $\alpha$ corresponds to an angular span of order $1/N$ hence a probability of $1/N$ of getting that sector on a throw. Thus, to hike up the probability of getting the jackpot to order unity, I must perform $O(N)$ iterations of the operations.

This result is hardly expected to please – after all this jazz I have taken the same number of steps as any person in the street would have used to find the jackpot by brute force ! And surely the $O(N)$ steps I have adopted are each individually very complex – invoking an ant army to sweep over all the sectors of the roulette and what not. This $O(N)$ answer is not a surprise though – it is well known that the fastest classical solution to this problem does indeed use $O(N)$ steps. It is now that we see the difference between classical and quantum mechanics. Firstly, the complex machinations performed by the ants' army have trivial quantum counterparts. Secondly and more importantly, in the quantum system we work with amplitudes and not probabilities. The starting value of these amplitudes (equivalent to the starting number of balls per sector) will be of the order of $1/N^{1/2}$ rather than $1/N$ as it was for the balls. Likewise the increment in amplitude at each step will be by an amount of $1/N^{1/2}$ and only $O(N^{1/2})$ iterations of the operations will be required to bring the jackpot amplitude up to order unity. This is the physical essence of Grover's searching algorithm. It also gives a reason why a quantum search algorithm should be able to perform the search in $O(N^{1/2})$ steps. Interestingly this same phenomenon also explains why quantum particles passed through a double slit (as in Feynman's Lectures, Volume 3) show an interference pattern rather than a doubly peaked distribution. These apparently unrelated effects both occur because quantum particles are described by amplitudes rather than probabilities.

In this last paragraph I briefly describe the quantum problem as it appears in the original paper and show the one-to-one correspondence between the quantum system and the roulette. I am given an $n$-qubit system which can be in a superposition of $N=2^n$ states (like the roulette). There exists an operator $C$ such that $C$ acting on exactly one of the $N$ states produces 1 (the jackpot), while $C$ on any other of the $N$-1 states gives zero. My task is to find the special state as quickly as possible. The first step is to create a state which is an equal superposition of all the $N$ possible states. This can be done by operating the matrix $\mathbf{M} = \dfrac{1}{\sqrt{2}}\begin{bmatrix} 1 & 1 \\ 1 & -1 \end{bmatrix}$ on each of the constituent qubits. By at most $O(\log N)$ iterations of $\mathbf{M}$ it is possible to bring the system to the superposition state where all the constituent amplitudes are equal and positive. For the roulette, this step is the equivalent of starting with a uniform distribution of all the balls. The next step is to iterate the two operations. The ball removal operation is implemented using a phase rotator – if the state $S$ of the system is such that $C(S)=1$ then the amplitude of the corresponding state is phase-shifted by $180^o$ i.e. reversed in sign. This causes this particular amplitude to go below the average amplitude (which is positive) and hence is the equivalent of removing the balls from the jackpot sector. Since the initial preparation is an equal superposition of all states, a single application of this phase rotation will have the desired effect on the target. Finally, the inversion about average is implemented through the diffusion transform operator $D = -\mathbf{I}(N) + \dfrac{1}{2N}\mathbf{ones}(N)$, where $\mathbf{I}$ is the identity matrix and $\mathbf{ones}$ denotes a matrix all whose entries are unity. It is readily shown that $(1/2)\mathbf{ones}(N)$ is a projection matrix which, acting on a vector $\mathbf{v}$, produces a new vector whose components are all equal to the average of the components of $\mathbf{v}$. It follows that $D$ denotes the inversion about average and is equivalent to the redistribution of the balls inside the roulette. I have already explained how the number of iterations is determined; a more rigorous derivation follows by letting the target state have the amplitude $(1-X^2)^{1/2}$ and all other states have amplitudes $X/N^{1/2}$. Now so long as $X^2 > 1/2$, the increase in amplitude at each iteration is at least $1/(2N)^{1/2}$. Hence in $N^{1/2}$ steps, the amplitude of the target state will be at least $1/2^{1/2}$. The final step (equivalent of throwing the roulette at the end) is to perform a measurement of the synthesized state; it will be in the target state with a probability of at least $1/2$.

\*     \*     \*     \*     \*

# ACKNOWLEDGEMENT

I am grateful to Kishore Vaigyanik Protsahan Yojana (KVPY), Government of India, for a generous Fellowship.